# On an innovative architecture for digital immunity passports[1] and vaccination certificates

John C. Polley[2], Ilias Politis[3], *Member, IEEE*, Christos Xenakis[4], *Member, IEEE*, Adarbad Master[5], and Michał Kępkowski[6]

*Abstract* — With the COVID-19 pandemic entering a second phase and vaccination strategies being applied by countries and governments worldwide, there is an increasing expectation by people to return to normal life. There is currently a debate about immunity passports, privacy, and the enablement of individuals to safely enter everyday social life, workplace, and travel. Such digital immunity passports and vaccination certificates should meet people's expectations for privacy while enabling them to present to 3rd party verifiers tamper-evident credentials. This paper provides a comprehensive answer to the technological, ethical and security challenges, by proposing an architecture that provides to individuals, employers, and government agencies, a digital, decentralized, portable, immutable, and non-refutable health status cryptographic proof. It can be used to evaluate the risk of allowing individuals to return to work, travel, and public life activities.

## I. INTRODUCTION

The urgency to contain the COVID-19 pandemic has led governments to implement various restrictions to day to day life. The development of several vaccines approved by the World Health Organization (WHO) [1], which are currently available to the public, assists in ending this societal health threat. While the number of people being vaccinated is increasing, the need for a carefully designed system of digital immunity passports is evident.

The adoption of a worldwide system that verifies people's COVID-19 status, including vaccination records, centers around the solution of digital immunity passports. This system would enable the holders of such passports to re-enter social life, work, and travel in safety. However, the nature of what information should be held on an immunity passport fuels debates among proponents of the solution and right-to-privacy advocates. Additionally, the technical challenges facing the successful integration of digital immunity passports in everyday life are yet unsolved. Most countries currently depend on paper-based systems in which people receive a test result or a vaccination record card with basic information on it. These paper records lack the security features that would render them trusted official certificates. Nothing prevents these documents from being lost or stolen, and thus, the opportunities for fraud are high [2].

In addition to the technical, legal, and societal issues, digital immunity passports are going to face possibly the most demanding scientific challenges. The development of new vaccines for COVID-19 and the beginning of general public vaccinations, signals the need for a revision on WHO recommendations for the COVID-19 Public Health Emergency of International Concern (PHEIC) [3], which would include the COVID-19 vaccination certificates. Therefore, it is evident that there is a need for a digital credentials proofing system, which would offer immutable proof of a vaccination status. In this study the focus is placed on providing an answer to the technological challenges that the application of digital immunity passports and vaccination certificates is facing, ensuring that such answers would also address the privacy and ethical concerns, which have been raised recently [7].

This paper is introducing a holistic solution for addressing the challenges that digital immunity passports and vaccination certificates are facing. The solution is a mobile service that maps a person's vetted identity and biometrics to the phone and then, cryptographically binds it with their COVID-19 test and vaccination records. The person can then prove their status by utilizing the credentials with a QR code, Bluetooth transfer or NFC single tap. The basis of the solution lies on the premise that users can have their identity vetted online with no personal contact, and thus, creating a digital identity that is bounded to their mobile phone. The proposed Digital Immunity Passport (DIPA) solution allows users to access test-laboratory or central government portals, utilizing their own account credentials. This solution also generates an access token, to be used by third parties to access the user's COVID-19 test results or vaccination records. These results and records are securely transferred to the mobile application and their anonymized version can be immutably logged. The information that is securely stored in the mobile application can be presented to verifiers upon user's biometric identification and consent.

[1] This work has been funded in part with Federal funds from the National Institutes of Health, Department of Health and Human Services, under Contract No. 75N91020C00035 and in part by the European Union's Horizon 2020 Stimulating innovation by means of cross-fertilisation of knowledge program under Grant 824015 (H2020-MSCA-RISE-2018-INCOGNITO) and the Grant 826404 (H2020-SC1-FA-DTS-2018-1-CUREX).
[2] John C. Polley at Systems Security Lab of the University of Piraeus, Greece (email: j.c.polley@ssl-unipi.gr).
[3] Ilias Politis at InQbit Innovations SRL., Bucharest, 041386 Romania and with the Systems Security Lab of the University of Piraeus, Greece (email: ilias.politis@inqbit.io, ipolitis@ssl-unipi.gr).
[4] Christos Xenakis at the Systems Security Lab of the University of Piraeus, Greece (email: xenakis@unipi.gr).
[5] Adabard Master at iCrypto, Inc., Santa Clara, CA 95054 U.S.A. (email: info@icrypto.com).
[6] Michał Kępkowski at Computing, Macquarie University, Sydney Australia (email: michal.kepkowski@students.mq.edu.au).

The rest of the paper is organized as follows. Section II presents the main challenges that need to be addressed before immunity passports can be widely adopted by society. The emphasis is placed on the technological challenges that currently limit the wide implementation of digital immunity passports. In Section III, the platforms stakeholders are introduced along with the overall description of the proposed DIPA solution. The key innovations are discussed in Section IV, while main components and functionalities are described in Section V. A description of the use case that is considered for this paper is presented in Section VI. Section VII summarizes the benefits and added value gained by national health care organizations and governments from incorporating the proposed platform into their workflows. Last, Section VIII concludes the paper.

## II. CHALLENGES FOR A SUCCESSFUL ADOPTION OF DIGITAL IMMUNITY PASSPORTS

Recently, the discussion on the issuing and applicability of digital immunity passports have raised concerns regarding the technological, scientific, and ethical problems such certificates are encountering [3]. Any plan for incorporating a digital immunity passport for daily use should offer completely anonymous and untraceable information to prevent being accessed without the consent of the user by 3$^{rd}$ party verifiers such as individuals, public health organizations and others. Furthermore, the digital immunity passports and the vaccination certificates should be in the heart of a holistic automated real-time policy enforcement strategy. This strategy could keep public health organization employees, first responders, individuals, and other stakeholders, aware of the current status and any policy updates. Such a strategic decision is expected to minimize the risk of exposure.

For a technological proposal to be adopted as a successful solution for addressing the portability, immutability, privacy and security aspects of the digital immunity passports and vaccination certificates, it should address the following challenges:

1. Individuals (users) should be able to remotely register in an easy and expeditious manner.
2. Users' identities should be automatically and securely verified leading to a seamless onboarding experience.
3. Users should be able to access the solution upon verification via state-of-the-art strong authentication (preferably biometric based).
4. Individuals must be able to prove their test result or vaccination status in their work environment, travel, etc., minimizing the probability of unintentionally endangering others.
5. The proof should be regularly updated to ensure that individuals are always presenting the most updated results, since earlier negative tests may not be indicative of their present health status.
6. The proof itself and the mechanism that produces it, needs to possess such properties that render it indisputably, unsusceptible to fraud, collusion, and misrepresentation.
7. The proposed solution must be seamlessly integrated to the COVID-19 testing ecosystem, eliminating the need for modifications and upgrades of the existing workflows required to produce immutable proofs.
8. The proof must be portable and easy to convey to any requesting authority.
9. Multiple proof types must be supported – positive or negative infection status, antibody presence, vaccination records, etc.
10. The proposed solution should not store any Personally Identifiable Information (PII).
11. The proposed solution should support an "anonymous mode" to enable use cases such as entering restaurants, cinemas, public venues, etc. without the user being required to reveal his identity to the verifier.
12. The proposed solution should be able to operate with machine verifiers (no human agent present).
13. The proposed solution supports use cases that require less than a second verification (mass-transit, stadiums, etc.).
14. The proposed solution should not require any specialized hardware. Leverage existing widely adopted smartphone technologies.
15. The proposed solution should support consumer and person to person use cases (food delivery, uber, entry to private residence, etc.)
16. Dynamic context must be collected at proof points to facilitate data collection and contact tracing.

## III. DIGITAL IMMUNITY PASSPORT SOLUTION

### A. Stakeholders and Entities

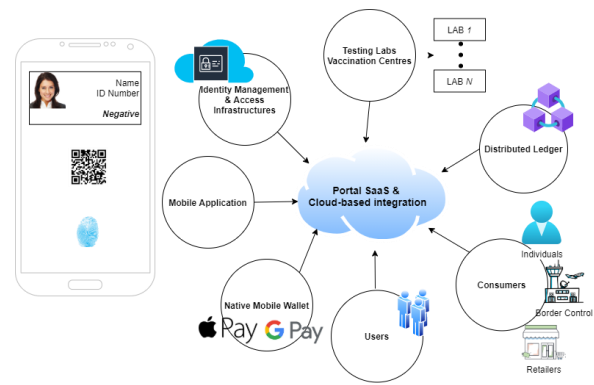

*Figure 1 Stakeholders and Entities*

The stakeholders and entities involved in the solution are depicted in Figure 1. The users of the mobile application are onboarded into the system and activated for usage of the service. The consumers of the service could include public and private organizations whose operation requires the immunity and vaccination status of the travelers, employees, etc. The testing laboratories and vaccination centers are also stakeholders of the proposed solution, which could convey the test results of the users to the application. The identity management and access infrastructure support the assignment of user identities and access privileges as well as the strong authentication of users. Moreover, the overall ecosystem displays the test and vaccination status and includes key entities, such as the mobile application responsible for interaction with the user to facilitate identity vetting, onboarding, and biometric authentication. The distributed ledger is utilized to create immutable audit trails and provide cryptographic proof of the immutability of the user verifiable credentials. The solution incorporates the integration of native mobile wallets to allow proofs to be in alternate portable formats, thus, ensuring quick validation based on existing widely adopted payment mechanisms. The solution is implemented as a software-as-a-service (SaaS) virtual private cloud to provide secure infrastructure services for building applications which store, process and share sensitive health-related information, in accordance with GDPR and other national security and privacy regulatory frameworks (i.e., HIPAA, etc.) [4]. The proposed solution does not store any PII of the user.

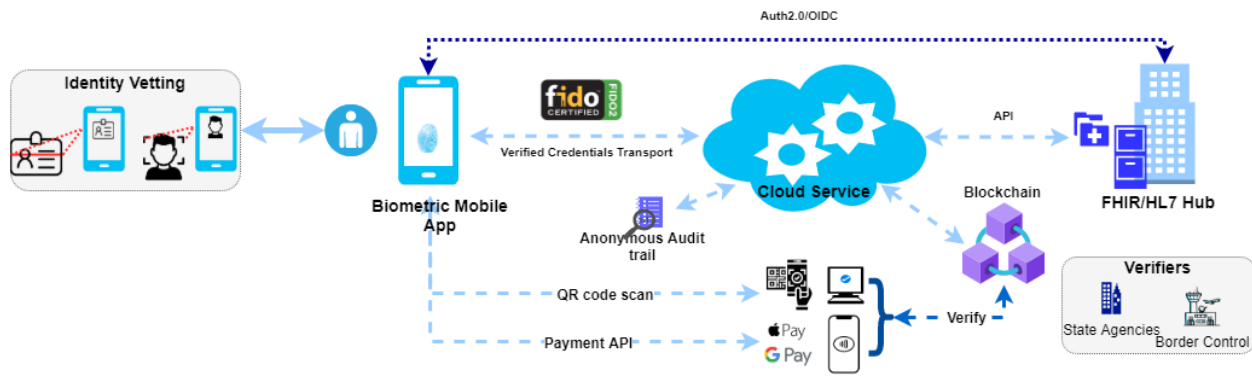

*Figure 2 Overview of DIPA proposed architecture*

*B. Overview*

An overview of the overall architecture, including the key protocols and actions described in the DIPA solution is shown in Figure 2. The solution allows a user to prove the status of their COVID-19 testing to provide verifiers (individuals, small business, authorities, etc.) with non-repudiated proof of vaccination records, current virus infection status and presence of antibodies to safely return to work, travel or socialize. Furthermore, the solution provides the user with the means to instantiate a trusted identity, using the current trusted institutional identities by leveraging a smartphone. Transport and application layer encryption along with state-of-the-art channel biding and certificate pinning techniques, establish secure transactions between the user and relevant stakeholders (i.e., test kit manufacturers, test laboratories, etc.).

The proposed technological solution ensures that the user's test status proof will remain resistant to fraud and cyberattacks, by incorporating a Keyless Signature Infrastructure (KSI) blockchain [8] (or any other Distributed Ledger technology), which record new events that are cryptographically linked to the previous ones in a distributed and immutable manner. Moreover, the solution allows the proof to be carried in a portable mobile wallet on the user's personal phone, by using security technologies such as, threshold cryptography and secure mobile storage. The smartphone's on-board fingerprint, face, and iris readers, in parallel to face recognition services, are utilized to ensure the user's identity proof and provide high degree of security. To minimize the need for modifications of the existing workflows of health care services and sufficiently address the ethical issues associated with the immutable proofs, the DIPA solution is designed to be fully compliant with the latest standards and regulations of health care, user privacy and data usage. One of the core principles in the designing and implementation of the proposed technological solution is that no personal and sensitive data of any kind is stored anywhere in the ecosystem, except the proof of status. Such data remains with the appropriate authoritative sources (i.e., vaccination centers, healthcare providers, government, and enterprise identity providers, etc.). The platform's seamless integration with current networks and services is ensured by leveraging widely established and well-known technologies/standards of identity, access management, trusted storage (OAuth2.0 [9] and Open ID Connect - OIDC [10], data storage {Federal Information Processing Standard – FIPS}, Transport Layer Security {TLS} [11], cryptographic proofs {hash functions} and distributed storage {KSI Blockchain}).

IV. KEY INNOVATIONS

Several key innovations have been introduced in the DIPA solution to ensure the digital immunity passport's integrity, the holder's privacy and the procedure's security.

*A. Digital onboarding*

The proposed solution allows users to leverage powerful aspects of technology such as mobility and privacy-preserving information sharing. The security in such environment begins with onboarding, that is the intuitive provisioning of individuals into the described ecosystem. Its implementation allows seamless onboarding of users, accompanied with assurances. Specific focus has been given on ensuring that there is no user manual data entry through this stage, thus, increasing its potential adoption by the community. Additionally, the digital onboarding allows the timely registration of biometrics, which will be used later, by the FIDO2 protocol [12]. A key advantage of the digital onboarding is the ability to perform online real-time identity verification, by exploiting the OCR and MRZ extractions along with performing checksum validations. The ID verification procedure is also enhanced with face matching between the ID and a selfie (with liveness detection - anti-spoofing), along with validating the ID data against the database of the identity issuer (i.e., government agencies, enterprise HR, etc.), in accordance with the use case and the national policies. Moreover, digital onboarding facilitates the creation of verifiable credential for the user identity signed by the identity management server and stored only on the user's phone. An immutable proof of all transactions is stored on the distributed ledger. The service has a built-in consent service, which capture user acceptance in a signed digital attestation.

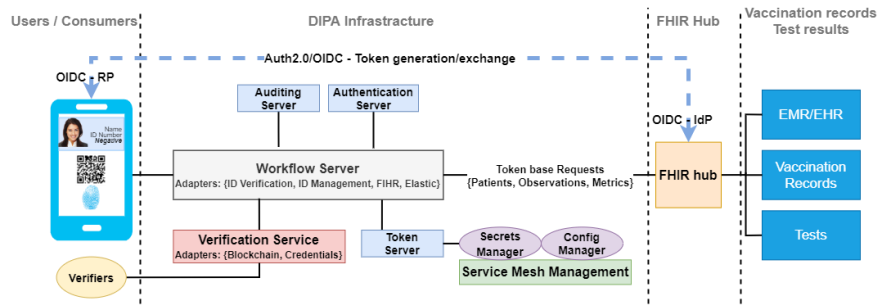

*Figure 3 DIPA key components and functionalities*

### B. Verifiable credentials

To address the immunity passport holders' demand for privacy and anonymity while assuring third party verifiers of the user's identity validity, the system design faces the following conundrum:
- Validate a user identity such that it is unquestionably associated with a person whose results are being requested from the health system.
- Do not keep any identifiable personal and health data on the solution servers.

DIPA is based on the implementation of the W3C Verifiable Credentials standard [14]. Essentially, the identity of the user and their status are represented as portable blocks of data that have incorporated cryptography mechanisms based on zero-knowledge proof. The generated credentials are sent to the mobile device over a secure channel. The mobile application then presents the credentials directly to any application (3rd party verifier) requesting the data.

Utilizing the verifiable credentials protocol, the system can store portable digital credentials on the mobile device only. Thus, it eliminates the need for keeping identity and test data on servers and allows the users to prove their identity and status directly to the requesting verifying party without any mediation of third-party servers. Furthermore, the user can select between identified or de-identified presentation of credentials. Such decision preserves the users' privacy and makes information disclosure a transaction directly between the holder (i.e., the user of the application) and the verifier (i.e., consumers), as in Figure 2. Firstly, the system ensures that the credentials themselves contain proofs of authenticity, since they are signed by the identity management server. Secondly, the system succeeds on guaranteeing the immutability of the verifiable credentials by deploying a distributed ledger that registers identifiers and user schemas. The verifiable credentials designed for the proposed system have a built-in expiration for protection against reuse, while they are specifically designed to be extensible enough to account for any credential claims. The scalability and easy incorporation of the solution to existing platforms is also supported by the wide interoperability of the W3C standard.

### C. Enhanced FIDO2

One of the challenges the DIPA solution is facing is the establishment of a cryptographically secure communications with the solution server that would allow the data to be sent with the biometric consent of the user. The selected solution to this challenge is the utilization of an enhanced FIDO2 based authentication. Specifically, the FIDO2 protocol is configured to utilize the extensions mechanism from WebAuthN [15]. The utilization of FIDO2/WebAuthN extensions allows the cryptographically signed data to be conveyed between application and server. Therefore, the data to and from the mobile device will be encapsulated within signed packets of binary objects. The signature will be generated via biometric authentication and will be verified by the server.

In parallel, the compliance to the standards facilitates an extensible architecture which can be easily enhanced with new verification methods and additional data elements. Finally, the extended FIDO2 creates a secure tunnel between mobile application and the authentication server with perfect forward secrecy.

### D. FHIR Interface

Instead of requesting from laboratories to integrate OIDC servers to provide users with the ability to allow authorization for third party applications to access their test results, the Health Level Seven International (HL7) standards body [16] has already proposed a standard which leverages OIDC/OAuth2 as an option for Web authentication, named Fast Healthcare Interoperability Resources (FHIR) [13]. The FHIR is a standard which describes data formats and elements, including an API for exchanging EHRs.

The implementation of FHIR within the proposed architecture offers an easier way to integrate mobile applications, such as the Digital Immunity Passport. Additionally, FHIR is ready for use and comes with out-of-the-box interoperability, thereby, reducing the Capital expenditures (CAPEX) and Operating expenses (OPEX). The supported user control over their data in FHIR forms the basis for the establishment of the user consent to share and self-report as envisioned in the DIPA solution. User consent is explicitly obtained when they request results from the vaccination centers or test labs. This consent is given online and is captured in digital format with all the context available to the DIPA system (i.e., geolocation of the mobile device, device data and time of day, etc.). The standard common target data format of the provided API ensures data integrity, accuracy, and consistency for a wide range of national health organizations data hubs, while it supports machine learning and artificial intelligent techniques for better data mining from these organizations.

*E. Workflow engine*

The workflow engine, incorporated into the design of the DIPA's technology framework, allows the integration of the implemented microservices into a scalable and distributed workflow engine. In parallel, it provides detailed visibility into how a workflow is performing so that it can identify potential problems. In addition, the workflow engine supports the orchestration of microservices to fulfill a defined workflow and ensures that all workflow instances are completed according to plan. Within this concept, it also allows the easy addition, update, and deletion of elements of a workflow, based on changes to user registration, FHIR data acquisition and verification flows. At the same time, it is capable to handle complex multi-stage registration and data acquisition workflows in scale with resilience and high performance.

## V. Main Components and Functionalities

The main architectural components and functionalities are illustrated in Figure 3. In detail, the mobile application is responsible first, for providing the user interface for the collection of the identity data. Secondly, for secure (encrypted) storage, the user's identity information as a verifiable credential and displaying the user, attributes whenever prompted. Moreover, it facilitates the communication with the FHIR hub to provide user access to the Electronic Medical or Health Records (EMR/EHR) portal, through the OAuth2/OIDC protocol, as well as, transporting the vaccination status or test results. The application is also charged with storing (encrypted) and displaying the test and the immunity status as verifiable credential.

The workflow server is offering web services to the mobile application. The role of this server is the provisioning of identity and biometric registration services using FIDO2/WebAuthN protocol and the facilitation of encrypted and signed transactions to/from the mobile device. The server is also providing various adapters for service operations, such as identity proofing and acquisition of vaccination status and test results. It handles the communications with the FHIR hub to register users as HL7 patient resource, while it provides observation interfaces that filter and log user and test result data, anonymously, on behalf of various National Health Organizations (NHO - e.g., NIH).

An identity management server is empowered with responsibilities associated with the processing and communication of the user's identity data. Specifically, this server receives image captures, optical character recognition (OCR) data and barcode readouts of identity documents and user selfie images from mobile application. Upon the reception of these data, it is extracting faces from the identity document and user selfies, submitting them to a face matching server. Furthermore, the server is used for capturing data such as name, date of birth and address from the ID document and deriving new identity data including city, state, age, etc. Finally, it is responsible for verifying ID documents with external authorities, creating identity verified credentials and inserting verification signatures into the distributed ledger. It is worth mentioning that no personal data is stored on any of the solution servers and after processing all sensitive data are discarded.

A token server is utilized to store and manage user tokens and keys. It is in control of storing push tokens for notification delivery to the mobile application and the public key management for the verification of the credential holder signatures from the mobile application. In addition, the token server is managing the symmetric keys for the QR code generation and verification.

The proposed architecture also incorporates an auditing server for providing high capacity and reliable indexed log storage services. The server is responsible for intaking audit trails for user registration and test results acquisition. It can retrieve data with text search and support raw and aggregated data visualization, while generating reports for data uploads to available data repositories of the various NHOs.

A verification service is designed for generating and validating verifiable credentials for users, test results and immutability statuses. Moreover, it is interfacing with the blockchain to create and verify transaction proofs. To offer a centralized service level configuration of all applications in the architecture, as well as a security module for storing and retrieving secrets (i.e., passwords, certificates, encryption keys, PKI, etc.) a specific component is defined in the architecture, namely the Service Mesh Management. It constitutes a set of infrastructure level services that provide core application and secrets management. It is designed to provide a policy engine for managing test results rationalization and access control.

Finally, the FHIR hub is envisioned as an external service, which interfaces with various EMR/EHR and proxies the HL7 interfaces to these entities. A wide section of EMR/EHR is compliant to the new HL7/FHIR standard, but this is not universal. As such, the component acts towards homogenizing the interface to EMR/EHR into a consistent set of REST APIs. In addition. it provides the required rationalization of HL7 observation resources into a consistent FHIR compliant JSON structure, while it can handle any security and compliance issues related to patient data acquisition.

As a result, the described architectural components constitute a holistic approach for allowing users to prove the status of their testing with respect to COVID-19, to provide 3rd party verifiers (individuals, enterprises, relevant authorities, etc.) with non-repudiated proof of current virus infection, presence of antibodies, vaccination records, and any other condition necessary to return to work, travel or socialize.

## VI. Use case

*A. Onboarding and Authentication*

As a first step, the user downloads and installs the mobile application associated with the proposed solution. Upon the completion of these two actions, the user is called to enable the device-based biometrics and is guided through an identity vetting process, which validates the identity of the user and cryptographically binds it to the device. DIPA guarantees that no PII data is stored on a server (privacy-preserving).

During this phase, DIPA registers user biometrics (PIN, fingerprint, face) on the phone using FIDO2. It then, requests the user to take a camera capture of the front and back of their government ID card. The user is then, requested to take a selfie of their face with liveness detection built in. The system verifies that the picture on the ID matches the selfie. All data from this process is conveyed to the backend where additional databases may be consulted for verification based on the ID type.

The result of this procedure is that a user identity verifiable credential is created. This is represented by a W3C Verifiable Credential with the associated decentralized identifier (DID) stored at a distributed ledger. This credential is portable and verifiable by any system that supports the appropriate standards. The credential is securely stored on the mobile. The user may also be asked to enter additional information such as email address. This information will be verified in the standard manner.

*B. Digital immunity test and vaccination certificate*

Although the procedure of obtaining a vaccination or COVID-19 test certificate may slightly differ per country, the user eventually is required to create an account at the specific EMR/EHR online portal, or similar systems provided by the various jurisdictions.

Using the mobile application, the user logs in to the EHR portal using a commercially available FHIR aggregation service. During this procedure, the login credentials are not exposed or stored neither in the mobile application, nor in the cloud-based servers. The FHIR hub provides an access token that is stored in the user's mobile phone secure storage. The mobile application will then call on the health system via the FHIR hub using FHIR protocols to retrieve the user's test results (i.e., observations). In this case, the access token is presented as a proof of authorization. The transaction is cryptographically signed and stored in an immutable audit trail using blockchain technology (i.e., immutable ledger). The proposed solution ensures that only the proof of status is stored on the distributed ledger – not the actual results data. The results, which constitute the actual data, are securely stored on the user's phone, and sent to the log database in anonymized format.

*C. Person-to-person fast presentation and verification of certificates*

The advantage of the existence of the mobile application is that it can hold the verifiable credentials (i.e., government ID, passenger location form, COVID-19 test, etc.) and utilize them in multiple ways depending on the use case. The solution support smartphone to smartphone verification without the need for specialized hardware. Upon entering the user biometrics, the application will cryptographically generate a QR code. Third party verifiers will be able to scan the QR code, which verifies in a tamper evident manner the user's test results on the immutable ledger.

Depending on the use case, DIPA supports different verification modalities with biometric authentication, verifiable credentials generation, cryptographic signature verification, and DL attestation. Moreover, to allow for seamless scalability and integration with an amalgam of communication modes and channels, DIPA supports dynamic and static QR code scan, NFC, Bluetooth, WiFi and mobile pass.

## VII. PLATFORM'S ADDED VALUE

The proposed Digital Immunity Passport provides a formal, provable, trust framework and ecosystem with integrated mobile applications, quick plug-in SDKs and componentized software microservices. These characteristics render this technology solution capable of rapid integration in operational healthcare facilities and COVID-19 testing services for issuing digitally signed credentials about a patient's COVID-19 status directly to their smartphones. With a quick biometric fingerprint or face scan, individuals can prove that they are currently virus-free or have been vaccinated. True to the spirit of verifiable credentials, this certificate does not contain personally identifiable information, hence, privacy and confidentiality can always be preserved while still providing strong cryptographic proof that the credential belongs to the specific individual. The proposed DIPA technology framework achieves:

- Peer to Peer tamper-evident COVID-19 test status and vaccination record verification.
- Privacy preserving, with credentials (government ID, test results, vaccination records) securely (encrypted) stored in the user's smartphone and only anonymized data for general purpose queries be allowed.
- Anonymous Verification by proving test results or vaccination status without shared ID information.
- Remote Onboarding with real-time ID verification by generating decentralized ID with Digital ID credentials stored on the smartphone.
- Biometric strong authentication with device biometrics.
- Extensible digital proofing framework based on the smartphone's secure credentials wallet.
- Future Proof Solution leveraging Industry open standards.

From the national healthcare organization's point of view, the DIPA solution presents the means for a secure drop-in solution deployable on-premises or in a private cloud as a SaaS. The numerous benefits that national health organizations, governments and others are gaining by incorporating the DIPA solution, include:

- Mapping the identity of the user to vaccination records and test results with an immutable digital certificate.
- Eliminating the uncertainty, fraud, repudiation, collusion, and impersonation by other individuals.
- Provision of security and frictionless experience easily adopted by users.
- Seamless integration into current workflows by test providers, laboratories, and enterprise systems.
- Compliance with privacy and consent collection hence, fostering user's trust in a seamless user experience.
- The implementation of mobile biometric and state-of-the-art open-source standards, which provides a modern and future-proof approach for test verification, scalable to new use cases.
- Digitally signed attestations with immutable audit logs, which provides compliance, accountability, and non-repudiation.

## VIII. Conclusion

As the world is experiencing the effects of the COVID-19 pandemic which affects almost every aspect of life (emotional, physical, financial, social), the debate about the introduction of an immunity passport that will allow a quicker return to a more familiar everyday life is heating up. Although privacy and ethical issues have been raised regarding the digital immunity passports, the existence of a digital certification for proof of COVID-19 status (i.e., testing, recovery, vaccination) may be necessary. The paper presents and discusses a framework solution for a scalable and commercially viable verifiable certificate augmented with non-repudiation and integrity. The proposed Digital Immunity Passport complies with the emerging worldwide regulations on privacy, trust, and accountability. It incorporates well studied and widely adopted by the industry standards (OAuth2/OIDC, FIDO2/WebAuthN, VC, FHIR/HL7, etc.) to empower national health care organizations and governments worldwide with the most robust technology for a next generation ubiquitous digital proofing solution, having a seamless access control and intuitive transactional verification.

## References

[1] https://extranet.who.int/pqweb/sites/default/files/documents/Status_COVID_VAX_16Feb2021.pdf . Accessed on 01 March 2021.

[2] Brown, Rebecca CH, Dominic Kelly, Dominic Wilkinson, and Julian Savulescu. "The scientific and ethical feasibility of immunity passports." The Lancet Infectious Diseases (2020).

[3] World Health Organization. "COVID 19 Public Health Emergency of International Concern (PHEIC). Global research and innovation forum: towards a research roadmap." (2020).

[4] Voigt, Paul, and Axel Von dem Bussche. "The eu general data protection regulation (gdpr)." A Practical Guide, 1st Ed., Cham: Springer International Publishing 10 (2017): 3152676.

[5] Annas, George J. "HIPAA regulations-a new era of medical-record privacy?." New England Journal of Medicine 348, no. 15 (2003): 1486-1490.

[6] Davahli, M.R.; Karwowski, W.; Sonmez, S.; Apostolopoulos, Y. The Hospitality Industry in the Face of the COVID-19 Pandemic: Current Topics and Research Methods. Int. J. Environ. Res. Public Health 2020, 17, 7366.

[7] Natalie Kofler, Françoise Baylis, "Ten reasons why immunity passports are a bad idea," Nature 581, 379-381 (2020)

[8] Nagasubramanian, Gayathri, Rakesh Kumar Sakthivel, Rizwan Patan, Amir H. Gandomi, Muthuramalingam Sankayya, and Balamurugan Balusamy. "Securing e-health records using keyless signature infrastructure blockchain technology in the cloud." Neural Computing and Applications 32, no. 3 (2020): 639-647.

[9] D. Hardt, "The oauth 2.0 authorization framework", IETF2012, 2012.

[10] Sakimura, Natsuhiko, John Bradley, Mike Jones, Breno De Medeiros, and Chuck Mortimore. "Openid connect core 1.0." The OpenID Foundation (2014): S3.

[11] Dierks, Tim, and Eric Rescorla. "The transport layer security (TLS) protocol version 1.2." (2008): 5246.

[12] FIDO-specs (2020). Fido specifications overview. https://fidoalliance.org/specifications.

[13] FHIR Overview, in: FHIR Release 4 (Technical Correction #1) (v4.0.1). Accessed October 12, 2020. http://www.hl7.org/fhir/overview.html.

[14] Manu Sporny et al., "Verifiable Credentials Data Model 1.0", W3C Rec, Nov 2019, [online] Available: https://www.w3.org/TR/vc-data-model/.

[15] Dirk Balfanz, et. al.,"Web Authentication: An API for accessing Public Key Credentials Level 1", W3C Recommendation, 4 March 2019

[16] Health Level Seven International – Homepage | HL7 International. Accessed October 12, 2020. https://www.hl7.org/.